\documentclass[a4paper,12pt]{article}
\usepackage{amssymb}
\usepackage{amsmath}
\usepackage{epsfig}
\usepackage{stmaryrd}
\usepackage{graphics}

\usepackage{latexsym}
\usepackage{rotating}

\title{\bf Direct search for exact solutions to the nonlinear Schr\"odinger equation}
\date{}

\author{
Wen-Xiu Ma$^{1}$\thanks{Email: {\tt mawx@cas.usf.edu}\ \ On
sabbatical leave of absence from University of South Florida, Tampa,
FL 33620, USA} \ and Min Chen$^{2}$
\\
{\small $^{1}$Department of Mathematics, Zhejiang Normal University,
Jinhua 321004, PR China}
\\{\small $^{2}$Department of Applied Mathematics, Donghua University, Shanghai 200051, PR China}
}

\begin{document}

\maketitle

\numberwithin{equation}{section}

\oddsidemargin=0mm \evensidemargin=0mm \baselineskip=17pt
\parindent 20pt

\newcommand \Z {\mathbb{Z}}
\newcommand \R {\mathbb{R}}
\newcommand \D {\displaystyle}

\def \part {\partial}
\def \be {\begin{equation}}
\def \ee {\end{equation}}
\def \bea {\begin{eqnarray}}
\def \eea {\end{eqnarray}}
\def \ba {\begin{array}}
\def \ea {\end{array}}
\def \si {\sigma}
\def \al {\alpha}
\def \la {\lambda}

\begin{abstract}
A five-dimensional symmetry algebra consisting of Lie point
symmetries is firstly computed for the nonlinear Schr\"odinger equation,
which, together with a reflection invariance, generates two
five-parameter solution groups. Three ans\"atze of transformations
are secondly analyzed and used to construct exact solutions to the nonlinear Schr\"odinger equation. Various examples of exact solutions with constant, trigonometric function type, exponential function type and rational function
amplitude are given upon careful analysis. A bifurcation phenomenon in the
nonlinear Schr\"odinger equation is clearly exhibited
during the solution process.

\vskip 2mm



\end{abstract}

\section{Introduction}

We are concerned with the cubic nonlinear Schr\"odinger (NLS)
equation
\begin{equation}
iu_t+u_{xx}+\mu |u|^2u=0,\label{eq:NLS:ma187}\end{equation}
where $u=u(x,t)$ is a complex-valued function of two real variables
$x,t$ and $\mu$ is a non-zero real parameter. The physical model of
the NLS equation \eqref{eq:NLS:ma187} and its generalized ones
occur in various areas of physics such as nonlinear optics, water
waves, plasma physics, quantum mechanics, superconductivity and
Bose-Einstein condensate theory
\cite{AblowitzS-book1981,HasegawaK-book1995}. In optics, the NLS
equation \eqref{eq:NLS:ma187} models many nonlinearity effects
in a fiber, including but not limited to self-phase modulation,
four-wave mixing, second harmonic generation, stimulated Raman
scattering, etc. For water waves, the NLS equation
\eqref{eq:NLS:ma187} describes the evolution of the envelope of
modulated nonlinear wave groups. All these physical phenomena can be
better understood with the help of exact solutions when they exist
for particular values of the parameter $\mu $.

It is well known that the NLS equation \eqref{eq:NLS:ma187} admits the bright soliton
solution \cite{ZakharovS-SPJETP1972}:
\begin{equation}
u(x,t)= k\sqrt{\frac {2}{\mu }}\, \, \textrm{sech} (k(x-2\alpha
t))\, e^{i[\alpha x-(\alpha ^2-k^2)t]}
,\label{eq:brightsolitonsolutionofNLS:ma187}\end{equation}
where $\alpha $ and $k$ are arbitrary real constants,
 for the self-focusing case $\mu > 0$,
and the dark soliton solution \cite{ZakharovS-SPJETP1973}:
\begin{equation}
u(x,t)= k\sqrt{-\frac {2}{\mu }}\, \tanh (k(x-2\alpha t))\,
e^{i[\alpha x-(\alpha ^2+2k^2) t]}
,\label{eq:darksolitonsolutionofNLS:ma187}\end{equation} where
$\alpha $ and $k$ are arbitrary real constants, for the de-focusing
case $\mu < 0$. These solutions are valid under the localized
traveling wave assumption. The $n$-soliton solutions to both the
self-focusing NLS equation and the de-focusing NLS equation can be
computed by the inverse scattering transform, the Darboux
transformation and the Hirota bilinear method (see, say,
\cite{NeugebauerM-PLA1984,PolyaninZ-book2004,Chen-book2006}). Moreover, ${\tilde u}(x,t)=u(ix,-t)$ offers a B\"acklund transformation between the self-focusing NLS equation and the de-focusing NLS equation.

A considerable amount of research has been devoted to the study of
exact solutions including traveling wave solutions of the NLS
equation (see, say,
\cite{ItsK-ANUSSRA1976}-\cite{Zhang-CNSNS2009}). Both numerical and
analytical methods have been used in dealing with the related
problems.
Generally, exact solutions to nonlinear equations are hard to come
by, but it is significantly important in mathematical physics to
find new ideals or approaches to discover solitary wave solutions of
nonlinear equations. Recently, several interesting studies have been
published to show that the NLS equation \eqref{eq:NLS:ma187}
has many new types of exact solutions (see, for instance,
\cite{Khuri-CSF2004}-\cite{Zhang-CNSNS2009}).

In this paper, we would like to present some direct search
approaches to exact solutions of the NLS equation
\eqref{eq:NLS:ma187} and construct its exact solutions over
some region of $\R ^2$, including analytical solutions on the whole
plane of $x$ and $t$. In what follows, on one hand, a
five-dimensional symmetry algebra is presented and used to generate
the corresponding five one-parameter solution groups. On the other
hand, three ans\"atze of transformations are analyzed, and various
examples of exact solutions with constant, trigonometric function type, exponential function type and rational function amplitude are calculated in
detail, covering many known exact solutions in the literature. The presented ans\"atze are direct but powerful,
particularly in getting traveling wave type solutions. A few
concluding remarks are given in the final section.


\section{Symmetry algebra and solution groups
}


We would like to present a five-dimensional symmetry algebra and
its corresponding one-parameter solution groups.

Obviously, the linearized equation of the NLS equation
\eqref{eq:NLS:ma187} is given by
\begin{equation} i\sigma _t+\sigma _{xx} +2\mu |u|^2\sigma +\mu u^2 \bar
\sigma=0,\label{eq:linearized equationofNLSeq:ma187}
\end{equation}
where $\bar \sigma$ is the complex conjugate of $\sigma$. It is
direct to check that there are five local Lie-point symmetries:
\begin{equation}\sigma_1=iu,\ \sigma_2=u_x,\ \sigma_3=u_t, \
\sigma_4=ixu-2tu_x,\ \sigma_5=u+xu_x+2tu_t,
\end{equation}
namely, the five functions $\sigma_i$, $1\le i\le 5$, satisfy the
linearized equation \eqref{eq:linearized equationofNLSeq:ma187}
when $u$ solves the NLS equation \eqref{eq:NLS:ma187}. These
are special reductions of symmetries of the general AKNS systems
\cite{Ma-JMP1992}.

Let us recall that the commutator of vector fields is defined by
\begin{equation} [K_1,K_2]=K_1'(u)[K_2]-K_2'(u)[K_1],\end{equation} where
$K'(u)[S]$ denotes the Gateaux derivative $K'(u)[S]=\frac {\partial
}{\partial \varepsilon}\bigl.\bigr|_{\varepsilon=0}K(u+\varepsilon
S)$.
 The symmetries $\sigma_i$, $1\le i\le 5$, constitute
a five-dimensional Lie algebra over the complex field under the
commutator of vector fields, and the non-zero commutators among
$[\sigma_r,\sigma_s]$, $1\le r< s\le 5$, are as follows:
\begin{equation} [\sigma_2,\sigma_4]=\sigma_1,\ [\sigma_2,\sigma_5]=\sigma_2,
\ [\sigma_3,\sigma_4]=-2\sigma_2,\ [\sigma_3,\sigma_5]=2\sigma_3,\
[\sigma_4,\sigma_5]=-\sigma_4.
\end{equation}
 The symmetries
$\sigma_1,\sigma _2,\sigma_3$ correspond to the $u$-scale
invariance, the $x$-translational invariance and the
$t$-translational invariance, respectively. The symmetries
$\sigma_4$ and $\sigma_5$ are two of so-called $\tau$-symmetries
\cite{Ma-JPA1990}, obtained from Galilean invariance and general
scale invariance. The general symmetry algebra of the NLS equation
\eqref{eq:NLS:ma187} contains Lie B\"acklund symmetries and
other but non-local $\tau$-symmetries.

It is evident to see that the symmetries $\sigma_i$, $1\le i\le 5$,
generate five one-parameter solution groups as follows:
\begin{equation} \left\{ \ba{l} \tilde u_1(x,t)=e^{i\varepsilon
}u(x,t),\vspace{2mm}\\
\tilde u_2(x,t)=u(x+\varepsilon ,t),\vspace{2mm}\\
 \tilde u_3(x,t)=u(x,t+\varepsilon ),\vspace{2mm}\\
 \tilde u_4(x,t)=e^{i(\varepsilon x-\varepsilon ^2t)}u(x-2\varepsilon
t,t),\vspace{2mm}\\
  \tilde u_5(x,t)=e^\varepsilon u (x e^\varepsilon, t e^{2\varepsilon}),
 \ea
\right.
\end{equation}
where $u$ solves the NLS equation \eqref{eq:NLS:ma187} and
$\varepsilon$ is a free real group parameter. Taking $\varepsilon
=\pi, \frac \pi 2 $, the first solution group $\tilde u_1$ yields
two special solutions $-u$ and $iu$, respectively.

Note that the NLS equation \eqref{eq:NLS:ma187} also has a
reflection symmetry. A coordinate reflection about the $t$-axis
generates a new solution $u(-x,t)$ from a known one $u(x,t)$.
Therefore, we can conclude the following two five-parameter solution
groups:
\begin{equation} \tilde u_\delta (x,t)=
 e^{i(\alpha xe^{\beta }-\alpha ^2te^{2\beta }+\eta_0)+\beta }u(\delta xe^{\beta }-2\delta \alpha te^{2\beta }
 +\xi_0 ,te^{2\beta }+\zeta_0 ),
\label{eq:BTofNLS:ma187}\end{equation} where $\delta =\pm 1$,
and the five free parameters $\eta_0,\xi_0,\zeta_0, \alpha,\beta $
correspond to the five symmetries $\sigma_1,\sigma_2,
\sigma_3,\sigma_4,\sigma_5$, respectively. These two solution groups
can be used to construct various new solutions from known ones.

\section{
Transformations and exact solutions}

We would, in this section, like to discuss three ans\"atze to
transform the NLS equation \eqref{eq:NLS:ma187} into real
simplified systems of differential equations and construct exact
solutions through those transformed NLS equations resulting from the
three ans\"atze.

\subsection{First ansatz}

We look for solutions by appending a phase factor to a real-valued
function. More precisely, we make an ansatz:
\begin{equation} u(x,t)=v(x,t)e^{i\eta },\ \eta =\alpha x+\gamma t,
\label{eq:1stAnsatz:ma187}\end{equation} where $v$ is a
real-valued function, and $\alpha,\gamma $ are two real constants.
This way, the NLS equation \eqref{eq:NLS:ma187} becomes a real
system:
\begin{equation}
v_t+2\alpha v_x=0,\ v_{xx}-(\gamma +\alpha ^2) v+\mu v^3=0.
\label{eq:TransformedNLS:ma187}\end{equation} If we set
$v=g/f$, then the system \eqref{eq:TransformedNLS:ma187} is put
into
\begin{equation}
\left\{ \ba {l} fg_t - f_tg +2\,\alpha\, f g_x -2\, \alpha\,
f_xg =0,
\vspace{2mm}\\
f^2g_{xx} -2 ff_xg_x +2 f_x^2g- ff_{xx}g -(\gamma + \alpha ^2) f^2
g +\mu\, g ^{3}=0.
 \ea \right. \end{equation} The first equation is bilinear while the second
one is trilinear. This setup brings us a direct approach to
searching for exact solutions. We can guess the type of functions
$f$ and $g$ first, and then look for exact solutions, especially by
computer algebra systems. For instance, a direct computation with
Maple can show that there is no multiple traveling wave type
solutions among the set of functions $u=(g/f)e^{i\eta }$ with
\[f=\sum_{m=0}^n a_m \, e^{k_mx + \omega_mt} ,\ g=\sum_{m=0}^n b_m \, e^{k_m x + \omega_m t}, \]
where $a_m,b_m,k_m,\omega_m\ 1\le m\le n,$ are real constants.

Let us remark that if we look for solutions $v=v(\xi)$ with $\xi
=k(x-2\alpha t)$, where $k\ne 0$ and $\alpha $ are real constants,
the transformed NLS equation \eqref{eq:TransformedNLS:ma187}
reduces to
\begin{equation}
k^2v_{\xi\xi}-(\gamma +\alpha ^2) v+\mu v^3=0.
\label{eq:FurtherTransformedNLS:ma187}\end{equation} This
equation is integrable, and it can be integrated as follows:
\begin{equation} \int_{v(\xi_0)}^{v(\xi) }\frac 1 {\sqrt{2(\gamma +\alpha ^2)v^2-\mu
v^4 +C}}\, dv=\frac {\sqrt{2}}{2k} (\xi -\xi_0),
 \end{equation}
 where $C$ and $\xi_0$ are arbitrary real constants.
The resulting solutions contain both elementary function and Jacobi
elliptic function solutions.

In what follows, we would like to find exact solutions of the NLS
equation \eqref{eq:NLS:ma187} with elementary function
amplitude.

\subsubsection{Solutions with constant amplitude}

 A constant solution $v=v_0$ of
\eqref{eq:FurtherTransformedNLS:ma187} leads to a class of
exact uniform solutions of the NLS equation
\eqref{eq:NLS:ma187}:
\begin{equation} u(x,t)= v_0 \, e^{i[\alpha x +(\mu {v_0}^2 -\alpha ^2)
t]} ,\label{eq:constantsolofTNLSundersa1:ma187}
\end{equation} where $v_0$ and $\alpha$ are arbitrary real
constants. This solution corresponds to a plane wave in water waves
and $\alpha $ corresponds to a simple shift of carrier-wave wave
number \cite{Peregrine-JAMSB1983}. The special case of
\eqref{eq:constantsolofTNLSundersa1:ma187} with $v_0=1$ was also
analyzed by other ans\"atze \cite{Khuri-CSF2004,Wazwaz-CSF2008}.

\subsubsection{Solutions with trigonometric function type amplitude}

Among the functions
\[ u(x,t)= (c\sec \xi +d \, \textrm{csc}\, \xi )\, e^{i\eta },\
\xi=k(x-2\alpha t),\ \eta =\alpha x +\gamma t, \] where
$c,d,k,\alpha,\gamma$ are real constants, we have the following
solutions with sec- and csc-function amplitude:
\begin{equation}
u(x,t)=k\sqrt{-\frac 2 {\mu }} \, (\sec \xi) \, e^{i\eta }, \
u(x,t)=k\sqrt{-\frac 2 {\mu }} \, (\csc \xi) \, e^{i\eta },
\label{eq:seccscsolu:ma187} \end{equation}
where $ \xi =k(x-2\alpha t)$, $ \eta =\alpha x -(\alpha ^2 +k^2)t$,
and $k$ and $\alpha $ are arbitrary real constants.

Among the functions
\[ u(x,t)= \frac {b_0+b_1\tan \xi +b_2\tan ^2 \xi }{a_0+a_1\tan \xi }\, e^{i\eta }, \
\xi=k(x-2\alpha t),\ \eta =\alpha x +\gamma t,\] where
$a_i,b_i,k,\alpha,\gamma$ are real constants, we have the following
three solutions with tan-function type amplitude. The first solution
is
\begin{equation}
u(x,t)=k\sqrt{-\frac 2 {\mu }}\, \frac {a_0\tan \xi
-a_1}{a_0+a_1\tan \xi } \, e^{i\eta },
\label{eq:tansolu:ma187}\end{equation} where $\xi =k(x-2\alpha
t)$, $ \eta =\alpha x - (\alpha ^2 -2k^2)t$, and $a_0,a_1,k,\alpha$
are all arbitrary real constants satisfying $a_0^2+a_1^2\ne 0$. This
solution contains exact solutions with tan- and cot-function
amplitude, which correspond to $a_1=0$ and $a_0=0$, respectively.
The second and third solutions are
\begin{equation}
u=\frac {k\sqrt{-2\mu }}{\mu }\,(\cot \xi +\tan \xi )\, e^{i\eta }
\label{eq:2ndtansolu:ma187} \end{equation}
with $\xi =k(x-2\alpha t)$ and $ \eta =\alpha x-(\alpha ^2+4k^2)t,$
and
\begin{equation}
u=\frac {k\sqrt{-2\mu }}{\mu }\,(-\cot \xi +\tan \xi )\, e^{i\eta }
\label{eq:3rdtansolu:ma187} \end{equation}
with $\xi =k(x-2\alpha t)$ and $\eta =\alpha x-(\alpha ^2-8k^2)t,$
where $k$ and $\alpha$ are arbitrary real constants. Those two solutions \eqref{eq:2ndtansolu:ma187} and
\eqref{eq:3rdtansolu:ma187} can be simplified to
 the solution \eqref{eq:seccscsolu:ma187} with csc-function amplitude and
the solution  \eqref{eq:tansolu:ma187} with $a_0=0$, respectively.

\subsubsection{Solutions with exponential function type amplitude}

If we focus on the set of functions $v=g/f$ with \[ f=\sum_{m=0}^n
a_m\, e^{m \xi } ,\ g=\sum_{m=0}^n b_m\, e^{m \xi },\ \xi
=k(x-2\alpha t),\] where $a_m,b_m,\ 1\le m\le n$, are real
constants, we obtain the following solutions with exponential
function type amplitude.

The non-constant solution of first order (i.e., $n=1$) is
\begin{equation}
v(x,t)= \frac k 2 \sqrt {-\frac 2 \mu } \, \frac { e^{\xi}-a_0}{
 e^{\xi} +a_0}
,\label{eq:1firstorderexpsolv:ma187}\end{equation} where $\xi =
k (x-2\alpha t)$ and $\gamma=-(\alpha ^2+\frac {k^2}2)$ in
\eqref{eq:FurtherTransformedNLS:ma187}. It thus follows that
the corresponding solution reads
\begin{equation}
u(x,t)=
\frac k 2 \sqrt {-\frac 2 \mu } \, \frac { e^{\xi}-a_0}{
 e^{\xi} +a_0}
 \, e^{i\eta },
\label{eq:1firstorderexpsolu:ma187}\end{equation} where $\xi =k
(x-2\alpha t)$, $ \eta =\alpha x -(\alpha ^2+\frac {k^2}2)t $, and
$a_0$, $k$ and $\alpha $ are arbitrary real constants. This solution
contains solutions with tanh- and coth-function amplitude, in
particular, the dark soliton solution. It also gives the solutions
presented in \cite{Zhang-CNSNS2009} since $\cos (2\tan ^{-1}
e^{\tilde \xi})=(1-e^{2\tilde \xi})/(1+e^{2\tilde \xi})$.

The first non-constant solution of second order (i.e., $n=2$) is
\begin{equation}
v(x,t)=\frac {8a_0b_1k^2 e^{\xi } }{8a_0^2k^2+\mu b_1^2 e^{2\xi} }
=\frac {8c k^2 e^{\xi } }{ 8 k^2+\mu c^2 e^{2\xi} }
,\label{eq:1secondorderexpsolv:ma187}\end{equation} where $\xi
= k (x-2\alpha t)$ and $\gamma =-\alpha ^2+k^2$ in
\eqref{eq:FurtherTransformedNLS:ma187}. It thus follows that
the corresponding solution reads
\begin{equation}
u(x,t)= \frac {8c k^2 e^{\xi } }{ 8 k^2+\mu c^2 e^{2\xi} }\,
e^{i\eta } , \label{eq:1secondorderexpsolu:ma187}\end{equation}
where $\xi =k (x-2\alpha t)$, $ \eta =\alpha x - (\alpha ^2-k^2)t $,
and $c$, $k$ and $\alpha $ are arbitrary real constants. This
solution contains a special solution with sech-function amplitude
including the bright soliton solution if $\mu>0$, and a special
solution with csch-function amplitude if $\mu<0$
\cite{Wazwaz-MCM2006}. It also suggests that we can show that the
self-focusing NLS equation \eqref{eq:NLS:ma187} does not have
any exact solution with csch-function amplitude.

The second non-constant solution of second order (i.e., $n=2$) is
\begin{eqnarray}
&& v(x,t)
 = \frac {k(-2 \sqrt{-2\mu}\, \mu + 4c k\mu \, e^{\xi
}+c^2k^2\sqrt{-2\mu}\, e^{2\xi} ) }{2\mu (2\mu +c^2k^2\, e^{2\xi} )
} ,\label{eq:2secondorderexpsolv:ma187}\end{eqnarray} where
$\xi = k (x-2\alpha t)$, $\gamma =-(\alpha ^2 +\frac {k^2}2)$ in
\eqref{eq:FurtherTransformedNLS:ma187} and $c=a_1/b_0$. It thus
follows that the corresponding solution reads
\begin{equation}
u(x,t)= \frac {k(-2 \sqrt{-2\mu}\, \mu + 4c k\mu \, e^{\xi
}+c^2k^2\sqrt{-2\mu}\, e^{2\xi} ) }{2\mu (2\mu +c^2k^2\, e^{2\xi} )
}
 \, e^{i\eta },
\label{eq:2secondorderexpsolu:ma187}\end{equation} where $\xi
=k (x-2\alpha t)$, $ \eta =\alpha x - (\alpha ^2+ \frac {k^2}2 )t $,
and $c$, $k$ and $\alpha $ are arbitrary real constants.

\subsubsection{Solutions with rational function amplitude}

Among the set of functions $v=g/f$ with
\[ f=\sum_{r,s=0}^2 a_{rs}x^rt^s,\ g=\sum_{r,s=0}^2 b_{rs}x^rt^s, \]
where $a_{rs},b_{rs}$ are real constants, we have the following
class of exact solutions with rational function amplitude:
\begin{equation}
v(x,t)={\frac { -2b_{{0,0}}\sqrt {-2\mu}\,\alpha -\sqrt {-2\mu}\, x
+2 \sqrt {-2\mu}\,\alpha t +2a_{{0,1}}
 }{4{b_{{0,0}}}^{2}{\mu \alpha}^{2}+4b_{{0,0}}\mu \alpha
x-8 b_{{0,0}} {\mu \alpha}^{2} t +\mu {x} ^{2}-4\mu \alpha
xt+4{\mu \alpha}^{ 2}{t}^{2}+2{a_{{0,1}}}^{2}}}
,\label{eq:rationalsolv:ma187}\end{equation} where
$b_{0,0},a_{0,1},\alpha$ are arbitrary constants and $\gamma=-\alpha
^2$ in \eqref{eq:TransformedNLS:ma187}.
 It thus follows that
the corresponding solution reads
\begin{equation}
u(x,t)=v(x,t)\, e^{i(\alpha x-\alpha ^2t)},\end{equation}
 with $v$ being defined by \eqref{eq:rationalsolv:ma187}.
 Taking $b_{0,0}=0$ leads to
 \begin{equation}
u(x,t)={\frac { -\sqrt {-2\mu}\,x+2\sqrt {-2\mu}\,\alpha
t+2a_{{0,1}} } {\mu {x}^{2}-4\mu \alpha xt+4\mu{\alpha}^{2}
{t}^{2}+2{a_{{0,1}}}^{2}}} \, e^{i(\alpha x-\alpha ^2t)} .
\end{equation} A further reduction with $a_{0,1}=0$ yields a special solution:
\[ u(x,t)=\sqrt{-\frac {2} \mu }\, \frac 1 {x-2\alpha t}\,
e^{i(\alpha x-\alpha ^2 t)},
\]
which, upon selecting $\mu =-2$ and $\alpha =-2\xi $, gives the
solution presented by a limiting process in \cite{BarranK-JPA1999}.

\subsection{Second ansatz}

We make the second ansatz
\begin{equation} u(x,t)
=[p(x,t)+i\,q(x,t)]\, e^{i\gamma t
},\label{eq:sa2ofNLS:ma187}\end{equation} where $p$ and $q$ are
two real-valued functions and $\gamma $ is a real constant. This
way, the NLS equation \eqref{eq:NLS:ma187} becomes
\begin{equation}
\left\{ \ba {l} p_t+q_{xx}-[\gamma -\mu (p^2+q^2)]q=0,\vspace{2mm}\\
 q_t-p_{xx}+[\gamma -\mu (p^2+q^2)]p=0.
\ea \right. \label{eq:TNLSundersa2:ma187}\end{equation} It can
be transformed into a system of two trilinear equations if taking
$p=g/f$ and $q=h/f$.
 A solution with constant amplitude is
\begin{equation}
u(x,t)=(c+ i\,d)\, e^{i\mu (c^2+d^2)t},
\end{equation}
where $c$ and $d$ are arbitrary real constants. Another special
solution with $p=c$ (a constant) is
\begin{equation}
u(x,t)=[c+i\,d \tan (\frac {\sqrt{-2\mu} \, d}{2} x+ cd\mu t )]\,
e^{i\mu (c^2 -d^2)t},
\end{equation}
where $c$ and $d$ are arbitrary real constants.

If we search for solutions with $q=0$, then $p=p(x)$ and so, the
transformed NLS equation \eqref{eq:TNLSundersa2:ma187} reduces
to
\[
p_{xx}-\gamma p +\mu p ^3=0. \] The other case with $p=0$ can be
transformed into this case by using the auto-B\"acklund
transformation $u\mapsto iu$. The first ansatz with $\alpha =0$
provides solutions for the above case. For example, the class of
solutions \eqref{eq:1firstorderexpsolu:ma187} with $\alpha =0$
leads to
\begin{equation} u(x,t)=
 \frac k 2 \sqrt {-\frac 2 \mu } \, \frac { e^{kx}-a_0}{
 e^{kx } +a_0}\, e^{-i\frac {k^2}2t}
 . \end{equation}
This further gives the solutions with the tanh- and coth-function
amplitude \cite{Wazwaz-MCM2006}:
\[u(x,t)=
 \frac k 2 \sqrt {-\frac 2 \mu } \tanh (\frac k2 x)\, e^{-i\frac {k^2}2t}, \ u(x,t)=
 \frac k 2 \sqrt {-\frac 2 \mu } \coth (\frac k2 x)\, e^{-i\frac {k^2}2t},\]
which correspond to $a_0=1$ and $a_0=-1$, respectively.

A special solution of the system \eqref{eq:TNLSundersa2:ma187}
with non-constant $p$ and $q$ is given by
\begin{equation}p(x,t)= \frac {a \, e ^{k(x-2\alpha t)} \, \cos \alpha x }{b+c\, e
^{2k(x-2\alpha t) } },\
 q(x,t)= \frac {a \, e ^{k(x-2\alpha t)} \, \sin \alpha x }{b+c\, e ^{2k(x-2\alpha t)
 }},
 \end{equation}
 where $\mu a^2=8k^2bc$ and $\gamma =-\alpha ^2+k^2$ in \eqref{eq:TNLSundersa2:ma187},
 and thus, a particular solution of the NLS equation \eqref{eq:NLS:ma187} reads
 \begin{equation}
 u(x,t)=[p(x,t)+iq(x,t)]\, e^{i(-\alpha^2 +k^2) t},\label{eq:1solofNLSundersa2:ma187}
 \end{equation}
 which includes two solutions ${\tilde u}(x,t)=u(ix,-t)$ with $u(x,t)$ defined by \eqref{eq:brightsolitonsolutionofNLS:ma187} and \eqref{eq:darksolitonsolutionofNLS:ma187}.
Other two solutions of the NLS equation \eqref{eq:NLS:ma187}
within this ansatz are
\begin{equation}
u(x,t)= \sqrt{\frac 2 \mu } \left[ \frac { 2ab^2\cosh (2a^2bc
t)+2iabc \sinh(2a^2bc t) }{ 2\cosh (2a^2bc t)\pm \sqrt{2}\,c \cos
(\sqrt{2}\, ab x) }-a \right] e^{2ia^2t},\ c=\sqrt{2-b^2}\,,
\label{eq:2solofNLSundersa2:ma187}
\end{equation}
where $a$ and $b\le \sqrt {2}$ are arbitrary real constants. Those
solutions were generated by the inverse scattering
transform in \cite{Ma-SAM1979}. Taking a limiting reduction of
$b\to 0$, the solution \eqref{eq:2solofNLSundersa2:ma187} with the minus sign gives an interesting solution
\cite{Peregrine-JAMSB1983,Debnath-book2000}: \[u(x,t)= \sqrt{\frac 2
\mu }\, \frac {3a-16a^5t^2-4a^3x^2+16ia^3t}{1+16a^4t^2+4a^2x^2}\,
e^{2ia^2t},\] a special case of which was also analyzed by using the
Adomian decomposition method \cite{El-SayedK-AMC2006}.

\subsection{Third ansatz}

We make the third ansatz
\begin{equation} u(x,t)=[p(x,t)+i\, q(x,t)]\, e^{i\alpha x },\label{eq:sa3ofNLS:ma187}\end{equation}
where $p$ and $q$ are two real-valued functions, and $\alpha $ is a
real constant. This way, the NLS equation \eqref{eq:NLS:ma187}
becomes
\begin{equation}
\left\{ \ba {l} p_t+2\alpha p_x+q_{xx}-[\alpha ^2-\mu (p^2+q^2)]q=0,\vspace{2mm}\\
 q_t+2\alpha q_x-p_{xx}+[\alpha ^2-\mu (p^2+q^2)]p=0.
\ea \right. \label{eq:TNLSundersa3:ma187}\end{equation} It also
can be transformed into a system of two trilinear equations if
taking $p=g/f$ and $q=h/f$.
A solution with constant amplitude is
\begin{equation}
u(x,t)=(c+i d)\, e^{i\sqrt{\mu (c^2+d^2)}\, x},
\label{eq:constantsolofTNLSundersa3:ma187}
\end{equation}
where $c$ and $d$ are arbitrary real constants.

If we search for solutions with $q=0$, then the transformed NLS
equation \eqref{eq:TNLSundersa3:ma187} reduces to
\begin{equation}
p_t+2\alpha p_x=0,\ p_{xx}-\alpha ^2 p + \mu p ^3=0.
\label{eq:simpliedTNLSundersa3:ma187}\end{equation} The other
case with $p=0$ can be put into this case by using the
auto-B\"acklund transformation $u\mapsto iu$. A special solution of
the NLS equation \eqref{eq:NLS:ma187} through
\eqref{eq:simpliedTNLSundersa3:ma187} is given by
\[
u=p(\xi )\,e^{i\alpha x},\ \xi=k (x -2\alpha t),\] if the function
$p$ solves
\[ k^2p_{\xi\xi}-\alpha ^2 p + \mu p ^3=0,
\]
where $k$ and $\alpha $ are real constants. The first ansatz with
$\gamma =0$ provides solutions for this case.

The following several special solutions of the system
\eqref{eq:TNLSundersa3:ma187} with non-constant $p$ and $q$ can
be obtained immediately, based on our analysis within the first
ansatz: \begin{equation} p=k\cos [(\mu k^2-\alpha ^2)t ],\ p=k\sin
[(\mu k^2-\alpha ^2)t ], \label{eq:sol1ofTNLSundersa3:ma187}
\end{equation} \begin{equation} p= k\sqrt
{-\frac {2}{\mu }} \,f(\xi) \cos [(\alpha ^2+k^2)t] ,\ q= -k\sqrt
{-\frac {2}{\mu }}\, f (\xi) \sin [(\alpha ^2+k^2)t] , \ f(\xi)=\sec
\xi \ \textrm{or}\ \csc \xi,
\end{equation}
\begin{equation} p= k\sqrt {-\frac {2}{\mu }}\, g( \xi) \cos [(\alpha ^2-2k^2)t] ,\
q= -k\sqrt {-\frac {2}{\mu }}\, g( \xi) \sin [(\alpha ^2-2k^2)t] ,\
g(\xi)=\tan \xi \ \textrm{or}\ -\cot(\xi),
\end{equation}
and
\begin{equation} p=
\frac {8ck^2 e^\xi }{8k^2 +\mu c^2 e^{2\xi}} \cos [(\alpha
^2-k^2)t] ,\ q= - \frac {8ck^2 e^\xi }{8k^2 +\mu c^2 e^{2\xi}} \sin
[(\alpha ^2-k^2)t] ,
\end{equation}
 where $\xi =k(x-2\alpha t)$, and $k$, $\alpha $ and $c$ are
arbitrary real constants.
 Another interesting solution of the NLS equation
 \eqref{eq:NLS:ma187} in
the form \eqref{eq:sa3ofNLS:ma187} with non-constant $p$ and
$q$ is given by
\begin{equation}
u(x,t)=(p+iq)\, e^{i \alpha x },\ p=\sqrt {\frac {\alpha ^2-2k^2}{\mu }}, \ q= k\sqrt{-\frac {2}{\mu
}} \tan (kx -d(t)) ,
\end{equation}
where $
 d(t)=[2k\alpha +k\sqrt{2(2k^2-\alpha ^2)}\,]\,
t,
$
and $k$ and $\alpha $ are arbitrary real constants satisfying
$2k^2\ge \alpha ^2$.

\section{Concluding remarks}

We have analyzed a five-dimensional symmetry algebra and three
ans\"atze of transformations for the NLS equation. Five one-parameter
solution groups are explicitly presented, and various exact solutions
with constant, trigonometric function type, exponential function
type and rational function amplitude are computed in detail,
covering many exact solutions generated in the literature. Our solution
analysis also provides clear information about a bifurcation
phenomenon between the self-focusing and the de-focusing NLS
equations. There are several solutions, like
 \eqref{eq:constantsolofTNLSundersa1:ma187},
\eqref{eq:1secondorderexpsolu:ma187} and
\eqref{eq:1solofNLSundersa2:ma187}, which works for both the
self-focusing case and the de-focusing case. Only a few solutions,
like \eqref{eq:2solofNLSundersa2:ma187} and
\eqref{eq:constantsolofTNLSundersa3:ma187},
 are valid for the self-focusing case, and most of the presented solutions
 are valid for the de-focusing case.

We remark that the first ansatz can be put into the second ansatz
[or the third ansatz], upon absorbing the phase factor $e^{i\alpha
x}$ [or $e^{i\gamma t}$] into the amplitude function $p+iq$ in
\eqref{eq:sa2ofNLS:ma187} [or \eqref{eq:sa3ofNLS:ma187}].
Therefore, all solutions in the first ansatz also provide solutions
within the second ansatz and the third ansatz. But the later two
ans\"atze are more general than the first one. Further applications
of the three ans\"atze and combinations of the three ans\"atze with
the solution groups in \eqref{eq:BTofNLS:ma187} can engender
other exact solutions to the NLS equation \eqref{eq:NLS:ma187}.
The three ans\"atze can also be used to construct exact solutions to
other nonlinear equations in mathematical physics, for example, the
Davey-Stewartson equation \cite{DaveyS-PRSA1974} and the generalized NLS equations \cite{ZhangXYC-RMP2009,ZhangYCX-preprint2009}.

As discussed for many typical integrable equations (see, for
example, the KdV equation \cite{Ma-PLA2002,MaY-TAMS2005}, the Toda
lattice \cite{MaM-PA2004,Ma-book2006}, the Boussinesq equation
\cite{LiMLZ-IP2007,MaHL-NA2008}, and the 2D Toda lattice
\cite{Ma-MPLB2008}), we will adopt the determinant techniques to
construct different kinds of exact solutions to the NLS equation in
a future publication. All the working methods of solutions,
including the inverse scattering transform
\cite{AblowitzS-book1981}, Darboux transformation
\cite{MatveevS-book1991}, the algebro-geometric method
\cite{BelokolosBEIM-book1994} and the Hirota method
\cite{Hirota-book2004} and the determinant techniques, help us
exploit the diversity of exact solutions of the NLS equation.


\vskip 2mm

\newpage
\footnotesize

{\leftline{\normalsize {\bf Acknowledgment}}}


{\small The work was supported in part by the Established Researcher
Grant of the University of South Florida and the CAS faculty
development grant of the University of South Florida.}


\begin{thebibliography}{99}


\bibitem{AblowitzS-book1981}
M. Ablowitz, H. Segur, Solitons and the Inverse Scattering
Transform, SIAM, Philadelphia, 1981.

\bibitem{HasegawaK-book1995}A. Hasegawa, Y. Kodama,
Solitons in Optical Communications,
Oxford University Press, New York, 1995.

\bibitem{ZakharovS-SPJETP1972}V.E. Zakharov, A.B. Shabat,
Exact theory of two-dimensional self-focusing and one-dimensional
self-modulation of waves in nonlinear media,
Soviet Physics JETP 34 (1972) 62--69 (translated from Zh. \`Eksper. Teoret. Fiz. 61 (1971) 
118--134).

 \bibitem{ZakharovS-SPJETP1973}V.E. Zakharov, A.B. Shabat,
 Interation between solitons in a stable medium, Soviet Physics JETP 37 (1973) 823--828
 (translated from Zh. \`Eksper. Teoret. Fiz. 64 (1973) 1627--1639).

\bibitem{NeugebauerM-PLA1984}G. Neugebauer, R. Meinel,
General $N$-soliton solution of the AKNS class on arbitrary
background, Phys. Lett. A 100 (1984) 467--470.

\bibitem{PolyaninZ-book2004}
A.D. Polyanin, V.F. Zaitsev, Handbook of Nonlinear Partial Differential Equations,
 Chapman $\&$ Hall/CRC, Boca Raton, 2004.

\bibitem{Chen-book2006} D.Y. Chen, Introduction to Solitons, Science Press, Beijing, 2006.

\bibitem{ItsK-ANUSSRA1976}
A.R. Its, V.P. Kotljarov, Explicit formulas for solutions of a
nonlinear Schr\"odinger equation, Dokl. Akad. Nauk Ukrain. SSR Ser.
A
1976 
965--968, 1051.

\bibitem{Ma-SAM1979}Y.C. Ma, The perturbed plane-wave solutions of the cubic Schr\"odinger equation, Stud. Appl. Math. 60 (1979) 43--58.

\bibitem{Peregrine-JAMSB1983}D.H. Peregrine, Water waves, nonlinear Schr\"odinger equations and their solutions,
J. Austral. Math. Soc. Ser. B 25 (1983) 16--43.

\bibitem{AkhmedievEK-TMP1987}
N.N. Akhmediev, V.M. Eleonski\v{\i}, N.E. Kulagin, First-order
exact solutions of the nonlinear Schr\"odinger equation, Theoret.
Math. Phys. 72 (1987) 
809--818 (translated from Teoret. Mat. Fiz. 72 (1987) 
183--196).

\bibitem{Olmedilla-PD1987}
E. Olmedilla, Multiple pole solutions of the nonlinear Schr\"odinger
equation, Physica D 25 (1987) 
330--346.

\bibitem{Hai-JPA1992}W.H. Hai, General soliton solutions of an $n$-dimensional nonlinear
Schr\"odinger equation, J. Phys. A: Math. Gen. 25 (1992) L515--L519.

\bibitem{Khuri-AMC1998}S.A. Khuri, A new approach to the cubic
Schr\"odinger equation: An application of the decomposition
technique, Appl. Math. Comput. 97 (1998) 251--254.

\bibitem{BarranK-JPA1999}
S. Barran, M. Kovalyov,
 A note on slowly decaying solutions of the defocusing nonlinear Schr\"odinger equation, J. Phys. A: Math. Gen.
  32 (1999) 6121--6125.

\bibitem{TovbisV-PD2000}
A. Tovbis, S. Venakides, The eigenvalue problem for the focusing
nonlinear Schr\"odinger equation: new solvable cases, Physica D 146
(2000) 150--164.

\bibitem{Demiray-AMC2003} H. Demiray, An analytical solution to the
dissipative nonlinear Schr\"odinger equation, Appl. Math. Comput.
145 (2003) 179--184.

\bibitem{AblowitzPT-book2004}M.J. Ablowitz, B. Prinari, A.D. Trubatch, Discrete and Continuous Nonlinear Schr\"odinger Systems, Cambridge University Press, Cambridge, 2004.

 \bibitem{Khuri-CSF2004}
S.A. Khuri, A complex tanh-function method applied to nonlinear
equations of Schr\"odinger type, Chaos Soliton. Fract. 20 (2004)
1037--1040.

\bibitem{El-SayedK-AMC2006}S.M. El-Sayed, D. Kaya,
A numerical solution and an exact explicit solution of the NLS
equation, Appl. Math. Comput. 172 (2006) 1315--1322.

\bibitem{Wazwaz-MCM2006}A.M. Wazwaz, Reliable analysis for nonlinear
Schr\"odinger equations with a cubic nonlinearity and a power law
nonlinearity, Math. Comput. Modelling 43 (2006) 178--184.

\bibitem{AktosunDv-IM2007}T. Aktosun, F. Demontis, C. van der Mee, Exact solutions to the focusing nonlinear Schr\"odinger equation, Inverse Probl. 23 (2007) 2171--2195.

\bibitem{Wazwaz-CSF2008}
A.M. Wazwaz, A study on linear and nonlinear Schrodinger equations
by the variational iteration method, Chaos Soliton. Fract. 37
(2008) 1136--1142.

\bibitem{OzisY-CSF2008}
T. \"Ozi\c{s}, A. Y{\i}ld{\i}r{\i}m, Reliable analysis for obtaining
exact soliton solutions
 of nonlinear Schr\"odinger (NLS) equation, Chaos Soliton. Fract. 38 (2008) 209--212.

 \bibitem{Zhang-CNSNS2009}
H.Q. Zhang, New exact complex travelling wave solutions to nonlinear
Schr\"odinger (NLS) equation,
 Commun. Nonlinear Sci. Numer. Simul. 14 (2009) 668--673.

\bibitem{Ma-JMP1992}W.X. Ma, Lax representations and Lax operator algebras of isospectral and nonisospectral hierarchies of evolution equations, J. Math. Phys. 33 (1992) 2464--2476.

\bibitem{Ma-JPA1990}
W. X. Ma, K-symmetries and t-symmetries of evolution equations and
their Lie algebras, J. Phys. A: Math. Gen. 23 (1990) 2707--2716.

\bibitem{Debnath-book2000}L. Debnath,
Nonlinear partial differential equations for scientists and
engineers (Second edition), Birkh\"auser Boston, Inc., Boston, MA,
2005.

\bibitem{DaveyS-PRSA1974}A. Davey, K. Stewartson, On three dimensional packets of surface waves, Proc. R. Soc. A 338 (1974) 101--110.

\bibitem{ZhangXYC-RMP2009}Y. Zhang, C.Z. Yao, X.N. Cai, H.X. Xu, A class of exact solutions of the generalized nonautonomous nonlinear Schr\"odinger equation, Rep. Math. Phys. 63 (2009) 427--439.

\bibitem{ZhangYCX-preprint2009} Y. Zhang, H.X. Xu, C.Z. Yao, X.N. Cai,  Wronskian solutions of a generalized nonautonomous nonlinear Schr\"odinger equation, preprint (2009).

\bibitem{Ma-PLA2002}W.X. Ma,
Complexiton solutions to the Korteweg-de Vries equation, Phys. Lett.
A 301 (2002) 35--44.

\bibitem{MaY-TAMS2005}W.X. Ma, Y.C. You,
Solving the Korteweg-de Vries equation by its bilinear form:
Wronskian solutions, Trans. Amer. Math. Soc. 357 (2005) 1753--1778.

\bibitem{MaM-PA2004} W.X. Ma, K. Maruno,
Complexiton solutions of the Toda lattice equation, Physica A 343
(2004) 219--237.

\bibitem{Ma-book2006}W.X. Ma, Mixed rational-soliton solutions to the Toda lattice
equation, in: Differential $\&$ difference equations and
applications, 711--720, Hindawi Publ. Corp., New York, 2006.

\bibitem{LiMLZ-IP2007}
C.X. Li, W.X. Ma, X.J. Liu, Y.B. Zeng, Wronskian solutions of the
Boussinesq equation---solitons, negatons, positons and complexitons,
Inverse Probl. 23 (2007) 279--296.

\bibitem{MaHL-NA2008}W.X. Ma, J.S. He, C.X. Li,
A second Wronskian formulation of the Boussinesq equation,
{Nonlinear Anal.} 70 (2009) 4245--4258.

\bibitem{Ma-MPLB2008}
W.X. Ma, An application of the Casoratian technique to the 2D Toda
lattice equation, Mod. Phys. Lett. B 22 (2008) 1815--1825.

\bibitem{MatveevS-book1991} V.B. Matveev, M.A. Salle, {Darboux Transformation
    and Solitons},
            Springer, Berlin, 1991.

\bibitem{BelokolosBEIM-book1994} E.D. Belokolos, A.I. Bobenko, V.Z. Enol'skii, A.R. Its,
V.B. Matveev, Algebro-Geometrical Approach to Nonlinear Integrable
Equations, Springer, Berlin, 1994.

\bibitem{Hirota-book2004} R. Hirota, {Direct Methods in Soliton Theory}, Springer,
    Berlin, 2004.

\end{thebibliography}
\end{document}